# Anisotropic Dzyaloshinskii-Moriya interaction and topological magnetism in two-dimensional magnets protected by $P\bar{4}m2$ crystal symmetry

Qirui Cui[†, ‡], Yingmei Zhu[†], Yonglong Ga[†], Jinghua Liang[†], Peng Li[†], Dongxing Yu[†], Ping Cui[†, ‡], Hongxin Yang[†, *]

[†]*Ningbo Institute of Materials Technology and Engineering, Chinese Academy of Sciences, Ningbo 315201, China; Center of Materials Science and Optoelectronics Engineering, University of Chinese Academy of Sciences, Beijing 100049, China*

[‡] *Faculty of Science and Engineering, University of Nottingham Ningbo China, Ningbo 315100, China*

*Corresponding author: hongxin.yang@nimte.ac.cn

**Abstract**

As a fundamental magnetic parameter, Dzyaloshinskii-Moriya interaction (DMI), has gained a great deal of attention in the last two decades due to its critical role in formation of magnetic skyrmions. Recent discoveries of two-dimensional (2D) van der Waals (vdW) magnets has also gained a great deal of attention due to appealing physical properties, such as gate tunability, flexibility and miniaturization. Intensive studies have shown that isotropic DMI stabilizes ferromagnetic (FM) topological spin textures in 2D magnets or their corresponding heterostructures. However, the investigation of anisotropic DMI and antiferromagnetic (AFM) topological spin configurations remains elusive. Here, we propose and demonstrate that a family of 2D magnets with $P\bar{4}m2$ symmetry-protected anisotropic DMI. More interestingly, various topological spin configurations, including FM/AFM antiskyrmion and AFM vortex-antivortex pair, emerge in this family. These results give a general method to design anisotropic DMI and pave the way towards topological magnetism in 2D materials using crystal symmetry.

KEYWORDS: two-dimensional magnets, crystal symmetry, Dzyaloshinskii-Moriya interaction, topological magnetism

Magnetic skyrmions are ideal information carriers for next-generation memory or logic technologies, such as racetrack memory[1,2], reconfigurable logic gates[3], artificial neuron devices[4], and quantum bit for quantum computing[5] with ultra-high density and low-energy consumption, thanks to their stable configuration with unique helicity, tunable nano-scale size, and low drive current density[6]. The key ingredient for the formation of skyrmion is DMI that originates from spin-orbit coupling (SOC) and inversion symmetric breaking (ISB)[7-10]. As a type of antisymmetric exchange coupling, DMI stabilizes preferred chirality of spin textures in acentric magnets. In the last decade, for achieving large interfacial DMI, much efforts have been denoted to growing ferromagnetic/heavy metal multilayers, where the spin orientations of transferred electrons between FM atoms are tilted due to the spin-orbit scattering of heavy atoms[11-13]. Recently, long-range magnetic order has been observed in 2D crystals[14-17] and gained great attention from both scientific and application field. Compared with bulk/multilayered magnets, these pristine 2D magnets possess much smaller size, simpler structure, higher tunability and better interface quality, which provide an intriguing platform for spintronic researches and applications[18]. We also note that the large DMI can be induced in ferromagnetic layers (Ni/Co bilayer) by chemisorbed light elements such as oxygen and hydrogen, and the chirality of DMI are further tuned by light elements coverage[19, 20]. These experimental results imply that sizable DMI is hopefully achieved in 2D magnets which even lacks heavy elements. Interestingly, experimental and theoretical studies have shown that FM skyrmion, bimeron, and chiral DWs can be established by isotropic DMI at 2D magnets-layered heterostructures[21-24] and 2D polar magnets with $C_{nv}$ point group, such as Janus and multiferroic monolayers[25-27]. However, the anisotropic DMI and AFM topological spin textures haven't been reported in pure 2D magnets so far. Anisotropic DMI favors the formation of topological magnetism, such as antiskyrmion which is a distinct type of skyrmion besides Bloch and Néel type, whose Hall angle can be directly controlled by the orientation of drive current[28-34]. In AFM systems, magnetic moments of coupled sublattices cancel out, resulting in zero dipolar field and enhances the stability of topological magnetism, at the same time, the topological charge of coupled sublattices also cancel out, resulting in zero skyrmion Hall angle and high mobility[35,36]. A very recent theoretical study shows that skyrmion Hall effects could not disappear for spin orbit torque-driven AFM skyrmion/skyrmionium due to the magnus force does not cancel of different skyrmion structure while spin Hall angle still vanishes for spin transfer torque-driven motion[37]. Notably, X-ray photoemission electron microscopy and spin-polarized scanning tunneling microscopy have been shown to unveil AFM spin configuration[38,39]. In a general perspective, anisotropic DMI-induced and AFM topological spin textures are highly promising for applications in advanced spintronic devices.

In this article, we show missing blocks for topological magnetism in 2D magnets. We propose the $AX_2$ monolayers with $P\bar{4}m2$ layer group for realizing crystal symmetry protected anisotropic DMI, where A is 3$d$ transition metal, and X is VI-A or VII-A element. Using first-principles calculations, we demonstrate that anisotropic DMI can be obtained in this family of 2D magnets, and due to the increasing of occupied 3$d$ electrons, FM phases are turned to be AFM phases when A varies from V to Ni. Moreover, using atomistic spin model, we reveal that various chiral spin configurations, including FM chiral DWs/antiskyrmion and AFM chiral DWs/antiskyrmion/vortex-antivortex pair, can be achieved in those $AX_2$ systems. The computational details are given in Supporting Information (SI) A-E.

Fig. 1(a)-(c) show the crystal structure of $AX_2$ monolayer. Each A atom is tetrahedrally surrounded by four X ligands. The layer group of $AX_2$ is $P\bar{4}m2$ which lacks inversion symmetry. From side views of $AX_2$ in Fig. 1(b) and (c), one can see that an X atom bonding with two A atoms along *x* and *y* directions are in bottom and top layer, respectively.

To investigate magnetic properties of our systems, we adopt following spin Hamiltonian:

$$H = -J_1 \sum_{<i,j>} \boldsymbol{S}_i \boldsymbol{S}_j - J_2 \sum_{<i',j'>} \boldsymbol{S}_{i'} \boldsymbol{S}_{j'} - K \sum_i (S_i^z)^2 - \sum_{<i,j>} \boldsymbol{D}_{ij} \cdot (\boldsymbol{S}_i \times \boldsymbol{S}_j), (1)$$

where $\boldsymbol{S}_i$ is unit vector indicating local spin of the *i*th A atom. $<i,j>$ and $<i',j'>$

represent summation of all the nearest-neighboring (NN) and next-nearest-neighboring (NNN) A pairs, and $J_1$ and $J_2$ represent the NN and NNN exchange coupling, respectively. $K$ refers to the magnetic anisotropy. We first discuss DMI vector in AX$_2$ monolayer, $\boldsymbol{D}_{ij}$, which is the key parameter in this work. $P\bar{4}m2$ layer group contains two mirror symmetries $M_x$ and $M_y$. According to Moriya symmetry (MS) rules[8], due to the existence of mirror plane across the middle of NN A-A bonds, the $\boldsymbol{D}_{ij}$ has the form as: $\boldsymbol{D}_{ij} = d_{\parallel}(\boldsymbol{u}_{ij} \times \boldsymbol{z}) + d_{ij,z}\boldsymbol{z}$, where $\boldsymbol{u}_{ij}$ is the unit vector between sites $i$ and $j$, and $\boldsymbol{z}$ is the out-of-plane (OOP) unit vector. A-X-A triplet coordinates in one mirror plane of AX$_2$, which results in that OOP component of DMI component $d_{ij,z}$ vanishes required by symmetry feature. Due to the rotoreflection $S_{4z}$ in $P\bar{4}m2$, the sign of in-plane (IP) DMI components should be opposite for $\boldsymbol{u}_{ij}$ along $x$ and $y$ direction ($\boldsymbol{u}_{ij,x}$ and $\boldsymbol{u}_{ij,y}$). Therefore, $\boldsymbol{D}_{ij}$ between the NN A atoms in $x(y)$ direction can be expressed as: $\boldsymbol{D}_{ij,x(y)} = d_{\parallel}^{x(y)}(\boldsymbol{u}_{ij,x(y)} \times \boldsymbol{z})$, where $d_{\parallel}^{x}=-d_{\parallel}^{y}$, thanks to the crystal symmetry protection. The orange arrows in Fig. 1(a) illustrate one possible configuration of $\boldsymbol{D}_{ij}$ of AX$_2$ monolayer. For both FM and AFM magnetic orders, this assumptive anisotropic DMI could favor anticlockwise (ACW) and clockwise (CW) spin spiral for an atomic chain along $x$ and $y$ direction, respectively. Another possible configuration of $\boldsymbol{D}_{ij}$ are shown in Fig. S1(a), which could favor ACW and CW spin spiral for an atomic chain along x and y direction, respectively.

The magnetism of most discovered 2D magnets, such as VSe$_2$, CrI$_3$, CrGeTe$_3$, MnSe$_2$ and Fe$_3$GeTe$_2$[14-17,40,41], originates from partially occupied 3$d$ orbitals. Accordingly, we choose A in AX$_2$ monolayer as magnetic elements V, Cr, Mn, Fe, Co, and Ni. Under tetrahedral crystal field, $d$ orbitals split into $d_{xy}$, $d_{xz}$, $d_{yz}$ orbitals ($t_2$) with high energy level and $d_{x^2-y^2}$, $d_{z^2}$ orbitals ($e$) with low energy level. Among 366 candidates containing 3$d$ transition metal (range from V to Ni) in the open Computational 2D Materials Database C2DB[42,43], we screen out 23 2D magnets satisfying AX$_2$ formula and structural conditions to realize anisotropic DMI. We first give the DFT results about basic structure and magnetic properties of $AX_2$ family as shown in Table SI. We adopt the sign convention that $J > 0$ ($J < 0$) indicates the FM (AFM) coupling, and $K > 0$ ($K < 0$) indicates OOP (IP) magnetic anisotropy. In most $AX_2$ monolayers, the magnitudes of $J_1$ are much larger than that of $J_2$. Interestingly, $J_1$ are always FM for V and Cr compounds while turn to be AFM for Mn, Fe, Co and Ni compounds (exception for $MnS_2$). In tetrahedral crystal, $t_2 \leftrightarrow p \leftrightarrow e$ superexchange allows FM coupling, however, the $t_2 \leftrightarrow t_2$ and $e \leftrightarrow e$ direct exchanges prefer AFM coupling, and the strength of $t_2 \leftrightarrow t_2$ is much larger than that of $e \leftrightarrow e$[44]. Therefore, when 3d orbitals are no less than half-filled, strong AFM exchange coupling emerges, which responds for the variation of $J_1$ of $AX_2$ monolayer. Similar transformation between FM and AFM phases is reported in zinc-blende binary transition metal compounds[44]. One can also see that Mn atoms in $MnX_2$ (X = Cl, Br, I) monolayers have the highest magnetic moment (close to 5 $\mu_B$) which gradually decreases for chemical elements on both sides of Mn in 3d transition metal row of the periodic table. The magnetic moment indicates that Mn has a high spin configuration ($S = 5/2$), and its decreasing in other transition metals quantitatively follow the total spin number of 3d orbitals. This trend of magnetic moment indicates that 3d electrons occupation in $AX_2$ basically follow the Hund's first rule[45]. For $AX_2$ with a certain A, distinct $d \leftrightarrow p$ orbitals hybridization controlled by X induces the variation of specific value of magnetic moment. The result that Mn magnetic moment in $MnS_2$ is less than $MnX_2$ (X = Cl, Br, I) about 1 $\mu_B$ implies that the 3d orbital is less than half-filled, which is responsible for the FM exchange coupling.

Fig. 2 shows the calculated DMI of $AX_2$ family from chirality-dependent energy difference (CDED) approach[10]. We adopt sign convention that $d_\parallel > 0$ ($d_\parallel < 0$) favors ACW (CW) spin configuration. Notably, the anisotropic DMI ($d_\parallel^x = -d_\parallel^y$) is obtained in all $AX_2$ monolayers, which is consistent with previous analysis. Interestingly, the strength of DMI in some systems reach to several meV, e.g., $VSe_2$ (2.24 meV), $FeS_2$ (-2.85 meV) and $NiI_2$ (6.99 meV), which are comparable to many state-of-art ferromagnetic/heavy metal multilayers[10-12]. Furthermore, to demonstrate the validity of results obtained from CDED approach, we perform qSO method where we consider SOC effects within first-order perturbation theory using self-consistent calculations to obtain $E(\boldsymbol{q})$[46,47]. Here, $E(\boldsymbol{q})$ represents the energy functional of spin spiral in AX$_2$ monolayers where $\boldsymbol{q}$ is spiral vector, and the DMI energy can be estimated by: $\Delta E_{DM}[\boldsymbol{q}] = (E[\boldsymbol{q}] - E[-\boldsymbol{q}])/2$[see derivations in SI D. We choose VO$_2$ and MnBr$_2$ as examples for FM and AFM phase, respectively. The latter calculations demonstrate the emergence of topological quasiparticle in these two systems. Fig. 3(a) shows asymmetric energy dispersion $E[\boldsymbol{q}]$ of a $\sqrt{2} \times \sqrt{2} \times 1$ VO$_2$ supercell along $\Gamma - M$ and $\Gamma - M'$ directions in 2D Brillouin zone (BZ). These two high-symmetry directions in reciprocal space correspond for directions of the NN V pairs in real space [see inset of Fig. 3(a)]. For $\boldsymbol{q}$ along $\Gamma - M$, ACW rotating spin spiral is favorable, while for $\boldsymbol{q}$ along $\Gamma - M'$, CW rotating spin spiral becomes more favorable, demonstrating the anisotropic feature of DMI. Moreover, with calculated $\Delta E_{DM}[\boldsymbol{q}]$ showing good linear dependence on $\boldsymbol{q}$ nearby the $\Gamma$ point [see Fig. 3(b)], $d_\parallel^x$ and $d_\parallel^y$ can be determined to be 1.51 and -1.51 meV, respectively. The computational approaches of $d_\parallel$ are given in SI D. Different from VO$_2$, the lowest energy appears at $M/M'$ point in MnBr$_2$ [see Fig. 3(c)] which corresponds to G-AFM phase. Interestingly, the chirality of preferred spin spiral keeps opposite in $\Gamma - M$ and $\Gamma - M'$ directions. From the linear fit of $\Delta E_{DM}[\boldsymbol{q}]$ nearby M point [see Fig. 3(d)], we obtain that $d_\parallel^x$ and $d_\parallel^y$ equal to -0.31 and 0.31 meV, respectively. These results indicate that the $q$SO method has good consistency with CDED approach about chirality and magnitude of DMI of AX$_2$ systems.

In order to further investigate the origin of DMI, we calculate the layer resolved SOC energy difference $\Delta E_{soc}$ between opposite chiral spin configurations. We find that $\Delta E_{soc}$ has opposite preferred chirality with spin rotating along *x* and *y* directions. Fig. 4 shows the $\Delta E_{soc}$ with spin rotating along *x* direction. For X atom with weak SOC, such as O, S and Cl, large part of contribution to DMI originates from A atom. When SOC strength of X enhances (X varies from Cl to I), the contribution to total

DMI from X keeps increasing, and in most cases, $X_{bot}$ gives more DMI contribution than $X_{top}$. This feature is consistent to Fert-levy model[9] where A-$X_{bot}$-A can be considered as a noncentrosymmetric triplet as shown in Fig. 1(b). We also notice that for systems with small DMI, such as $FeO_2$ and $NiBr_2$, $\Delta E_{soc}$ from A layer and X layer has similar magnitude but opposite chirality, resulting in these contributions cancelling each other.

For $AX_2$ system, there are two two-fold rotation axes ($C_2$) passing through NNN A atoms and perpendicular to each other [see Fig. S1(b)]. Based on MS rules[8], DM vectors between the NNN magnetic atoms should be along directions connecting these atoms [indicated by red arrows in Fig. S1(b)], which favor helicoid spin propagations with opposite chirality [see Fig. S1(c)]. However, the magnitude of NNN DMI ($d_{NNN}$) is very tiny compared with the NN DMI. For example, using four-state energy mapping method[48], we find that $d_{NNN}$ of $VO_2$ is only 0.041 meV. Moreover, calculated spin textures (see latter discussions) show that despite the NNN DMI is neglected, the antiskyrmion with both helicoid and cycloid spin spiral appears in $VO_2$.

Once all magnetic parameters in spin Hamiltonian $H$ are obtained, one can define the effective magnetic field $\boldsymbol{B}^i_{eff} = -\frac{1}{\mu_s}\frac{\partial H}{\partial \boldsymbol{S}_i}$ and perform time evolution of spin on each atomic site to investigate real-space spin configurations, using the Landau-Lifshitz-Gilbert (LLG) equation: $\frac{\partial \boldsymbol{S}_i}{\partial t} = -\frac{\gamma}{(1+\lambda^2)}[\boldsymbol{S}_i \times \boldsymbol{B}^i_{eff} + \lambda \boldsymbol{S}_i \times (\boldsymbol{S}_i \times \boldsymbol{B}^i_{eff})]$, where $\gamma$ indicates gyromagnetic ratio and is set to $1.76 \times 10^{11}$ $T^{-1}s^{-1}$, and $\lambda$ represents damping constant and is set to 0.2. In $VO_2$ monolayer, FM antiskyrmion with diameter of 10 nm [yellow dashed square in Fig. 5(a)] emerges. There are Néel walls appearing in *x* and *y* directions connecting with four Bloch lines; and IP magnetic moments are along radial direction of these Bloch lines[30] [see Fig. 5(b)]. Using the formula $Q = \frac{1}{4\pi}\int \boldsymbol{S} \cdot (\partial_x \boldsymbol{S} \times \partial_y \boldsymbol{S})dxdy$, the calculated magnetic topological charge $Q$ is -1 for FM antiskyrmion in yellow dashed square. Due to the enhancement of FM exchange coupling and PMA, large size of domain separated by Néel DW appears in

VS$_2$ and VSe$_2$ monolayers [see Fig. S2(a) and (b)].

In range from CrO$_2$ to MnCl$_2$ in Table SI, uniform FM and G-AFM phases are observed, which is a consequence of much stronger Heisenberg exchange coupling compared with DMI. Interestingly, an isolated AFM antiskyrmion emerges on the background of large-size domain in MnBr$_2$ [see Fig. 5(c)]. In the zoomed AFM antiskyrmion shown in Fig. 5(d), one can see that there are two opposite spin-dependent sublattices with AFM coupling and opposite $Q$ (+1, -1). Despite $Q$ for AFM antiskyrmion is zero, it is topologically protected since it cannot be destroyed or split into pieces. Similar to skyrmionium, these spin configurations, with both vanishing topological charge and topological protection, could be created continuously from uniform magnetic background and move steadily without detrimental skyrmion Hall effect[49]. The diameter of this AFM antiskyrmion is only 6.8 nm, which can significantly enhance integration level of corresponding devices. There is an angle of 45° between orientations of Néel walls in FM and AFM antiskyrmion [see Fig. 5(b) and (d)]. This angle actually arises from the difference of chosen unit cell for FM systems ($1 \times 1 \times 1$) and AFM systems ($\sqrt{2} \times \sqrt{2} \times 1$) in atomistic spin model simulations. Compared with MnBr$_2$, the symmetric exchange coupling slightly decreases while DMI largely increases in MnI$_2$, which results in that AFM antiskyrmion is embedded in meandering domain [see Fig. S2(c)]. Notably, we observe the AFM vortex-antivortex pair in CoI$_2$ monolayer [see Fig. 5(e) and (f)], which is demonstrated by the $Q$ density [see later discussions]. Similar to AFM antiskyrmion, $Q$ of vortex or antivortex is zero but these spin textures are still topologically protected. For Ni-based systems, such as NiI$_2$, high-density wormlike domains separated by chiral DWs are achieved due to significant DMI strength [see Fig. S2(h)]. Above scenarios clearly show that various FM/AFM topological spin configurations are achieved in the family of AX$_2$. We also implement Monte Carlo (MC) metropolis algorithm for investigating the ground spin state of exemplary systems, VO$_2$, MnBr$_2$ and CoI$_2$ monolayers [see Fig. S3], and detailed discussion is given in the SI E.

The $Q$ density is further analyzed with the formula $\mathbf{S} \cdot (\partial_x \mathbf{S} \times \partial_y \mathbf{S})$, and for AFM systems, $Q$ density is calculated for two sublattices respectively. As shown in the Fig. S4, $Q$ distribution clearly corresponds to the morphology and location of topological quasiparticle. For FM antiskyrmion in VO$_2$ monolayer, the negative $Q$ is radical from center, while for AFM antiskyrmion in MnBr$_2$ monolayer, the negative and positive $Q$ of two sublattices are coupled together resulting in vanishing topological charge. In CoI$_2$ monolayer, the $Q$ distribution is around the cores of vortex and antivortex respectively, indicating that there is vortex-antivortex pair rather than bimeron soliton[27, 50, 51]. The later requires that $Q$ distribution of vortex and antivortex can't be separated. Moreover, we demonstrate the dynamic stability of systems with chiral magnetism monolayer via phonon dispersions [see Fig. S5]. Despite there is small acoustic imaginary mode nearby $\Gamma$ point for VS$_2$ and CoBr$_2$ monolayers, this instability may not significantly affect the whole crystal structure and can be removed by ripples in structures. Experiments have shown that using molecular beam epitaxy (MBE), tetrahedral crystal in thin film form can be grown on the appropriate substrate[52, 53].

The DMI/NN exchange coupling ratios $|d_\parallel / J_1|$ of all AX$_2$ monolayers are calculated [see blue dots of Fig. S6]. One can see that $|d_\parallel/J_1|$ of VO$_2$, FeI$_2$ and CoI$_2$ are 0.159, 0.192 and 0.118, respectively, which are in the typical range of 0.1-0.2 for the skyrmion formation[2]. Since spin textures are determined by $J$, $K$ and $d_\parallel$ cooperatively, and $K$ is neglected in $|d_\parallel/J_1|$, it is convenient to introduce another dimensionless parameter, $\kappa = |(\frac{4}{\pi})^2 \frac{2J_1 K}{d_\parallel^2}|$, for systems with magnetic atom arranging in 2D square lattice[22]. When $0 < \kappa < 1$, the magnetic ground state of system exhibits spin spiral, whereas for $\kappa > 1$, the isolated skyrmion can be generated[54, 55]. The results for AX$_2$ are indicated by red dots in Fig. S6. Contrary to $d_\parallel$ favoring spin spiral, strong $J_1$ and $K$ make spins coupled linearly, resulting in uniform FM or G-AFM phases. Therefore, chiral spin textures are difficult to emerge in AX$_2$ combining small $|d_\parallel/J_1|$ ($\ll 0.1$) and large $\kappa$ ($\gg 1$), such as CrX$_2$ (X=O, S, Se, I) and MnX$_2$ (X=S, Cl), but tend to appear in systems with large $|d_\parallel/J_1|$ and small $\kappa$, such as VO$_2$, AI$_2$ (A=Fe,

Co) and NiX$_2$ (X=Cl, I). Furthermore, we distinguish systems that hold chiral magnetism with light purple background, which are demonstrated by spin atomic model simulations. Despite only $J_1$ is included, the two parameters above can be considered as criteria for possible formation of spin textures since the magnitude of $J_2$ is very small in most cases.

In summary, we propose and demonstrate the anisotropic DMI in AX$_2$ monolayer with $P\bar{4}m2$ crystal symmetry protection. We further unveil that various FM/AFM topological quasiparticles can be stabilized without external field and provide two criteria that can predict possible spin configurations in these 2D magnets. Moreover, the structure of AX$_2$ can be considered as a crucial framework to search more 2D magnets that possess anisotropic DMI and non-trivial spin configurations in other material databases. Such materials have many unique advantages, such as flexibility, gate tunability, and miniaturization, to replace the complex bulk magnets or ferromagnet/heavy metal multilayers traditionally used to achieve topological magnetism and lead to development of spintronic devices with simpler structure and higher efficiency. Our work thus provides a robust route to construct crystal symmetry protected anisotropic DMI and topological magnetism in 2D magnets, which will be highly promising for future spintronic applications.

**Supporting Information**

(1) Computational details: A. The first-principles calculations, B. The out-of-plane polarization, C. Phonon calculations, D. Calculations of $J$, $K$, and $d_{\parallel}$, E. The atomic spin model simulations; (2) Schematic of the NN and NNN DMI; Real-space spin configurations VS$_2$, VSe$_2$, MnI$_2$, FeI$_2$, CoCl$_2$, CoBr$_2$, NiCl$_2$ and NiI$_2$; (3) Ground spin states and (4) topological charge density of VO$_2$, MnBr$_2$ and CoI$_2$; (5) Phonon dispersions; (6) Two criteria that could be used to predict spin textures; (7) Spin configurations used in calculations of $J$, which includes Ref. 24, 42, 43, 48, 56-71

**Acknowledgement**

This work was supported by the National Natural Science Foundation of China, Grants No.11874059 and No.12174405; Key Research Program of Frontier Sciences, CAS, Grant No. ZDBS-LY-7021; Zhejiang Provincial Natural Science Foundation, Grant No. LR19A040002; and Beijing National Laboratory for Condensed Matter Physics.

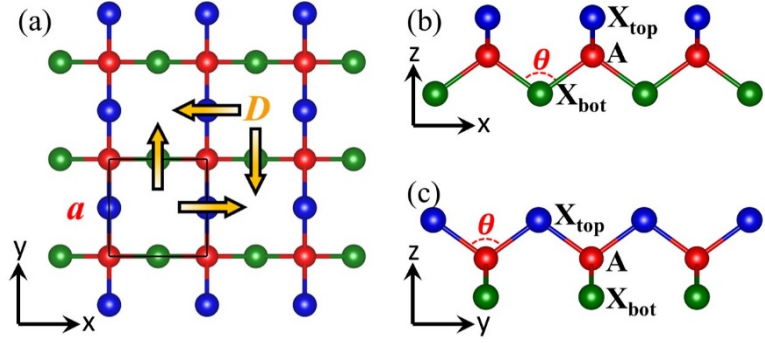

**Figure 1.** Geometric structures of $AX_2$ monolayer. Top view (**a**) and side views (**b**, **c**) of $AX_2$ monolayer. Red and blue/green balls represent magnetic and nonmagnetic elements respectively. The inset orange arrows of top view indicate the orientation of in-plane component of DMI between nearest-neighboring (NN) magnetic atoms. This DMI could favor the anticlockwise (ACW) and clockwise (CW) spin spiral along a chain of atoms in x and y direction, respectively.

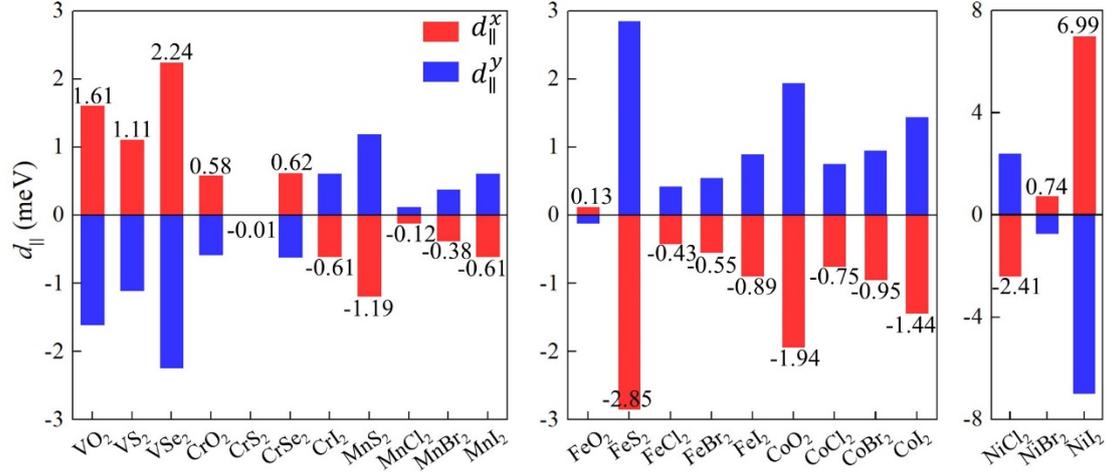

**Figure 2.** The calculated in-plane component of DMI for 23 AX$_2$ monolayers. Red and blue bars represent the DMI between NN magnetic atoms at *x* and *y* directions respectively. Here $d_\parallel > 0$ ($d_\parallel < 0$) favors spin canting with ACW (CW) chirality. It is noted that $d_\parallel^x = -d_\parallel^y$ in all systems, which demonstrates the anisotropic feature of DMI in AX$_2$. We give specific values of $d_\parallel^x$ as shown by numbers of red bars.

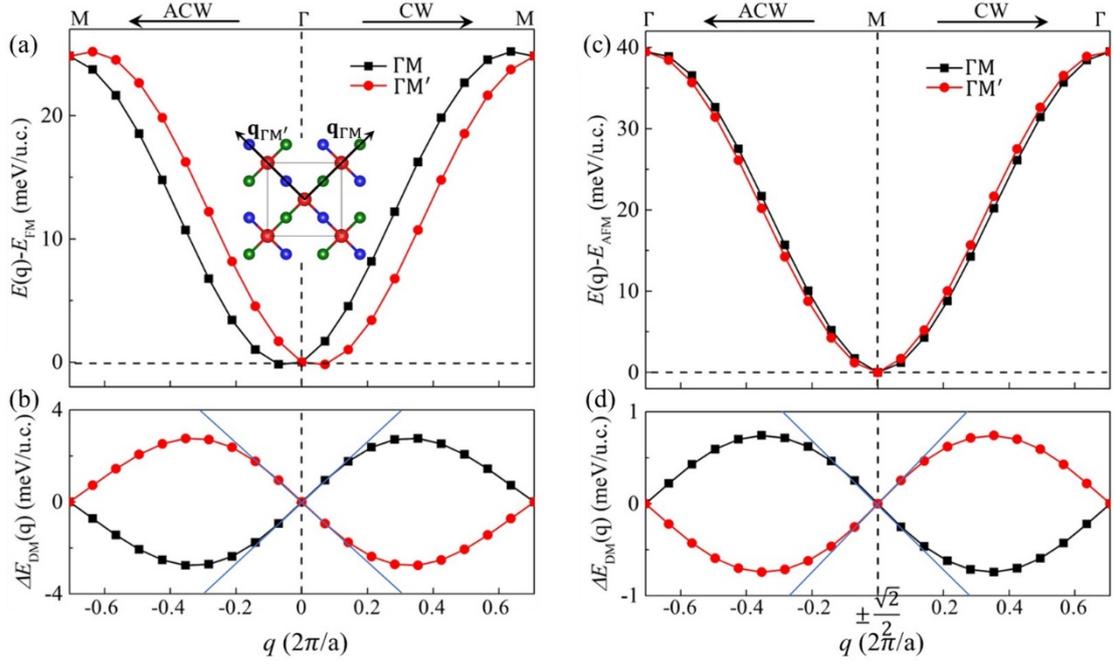

**Figure 3.** Calculation of magnetic parameters by qSO method. Spin spiral energy $E(q)$ (upper panel) and DMI energy $\Delta E_{DM}(q)$ (lower panel) as functions of spiral vector length $q$ for $VO_2$ (**a**, **b**) and $MnBr_2$ (**c**, **d**). In $VO_2$, $E(q)$ is given respect to the ferromagnetic state at $q = 0$ while in $MnBr_2$, $E(q)$ is given respect to the antiferromagnetic state at $q = \pm\sqrt{2}/2$. Black and red points are calculated with $\boldsymbol{q}$ along $\Gamma - M$ and $\Gamma - M'$ respectively (see inset of a). Blue lines in (**b**) and (**d**) are linear fits of $\Delta E_{DM}(q)$, which is based on the atomistic extended spin Hamiltonian.

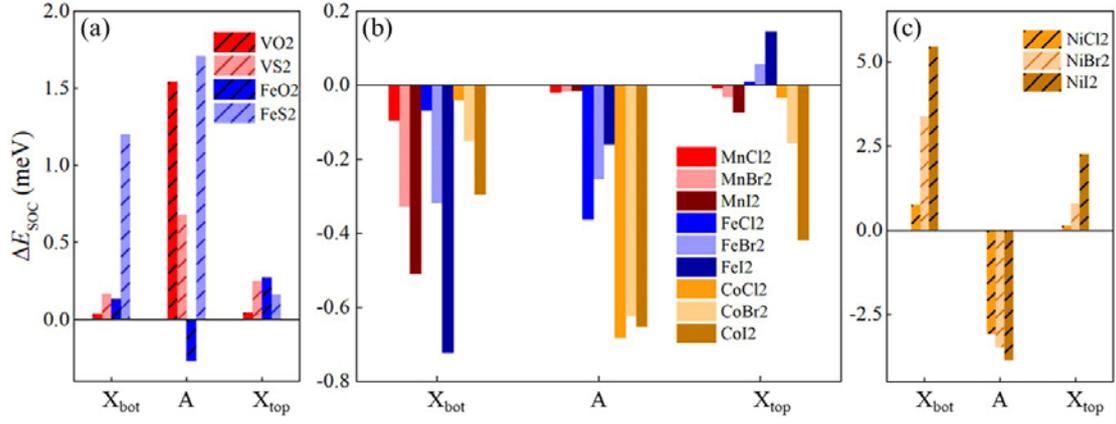

**Figure 4.** Anatomy of DMI for AX$_2$ monolayers. Atomic-layer-resolved localization of SOC energy difference $\Delta E_{soc}$ between opposite chiral spin configurations for vanadium/iron dichalcogenides (**a**), manganese/iron/cobalt dihalides (**b**) and nickel dihalides (**c**). Here we show $\Delta E_{soc}$ with spin rotating along *x* direction.

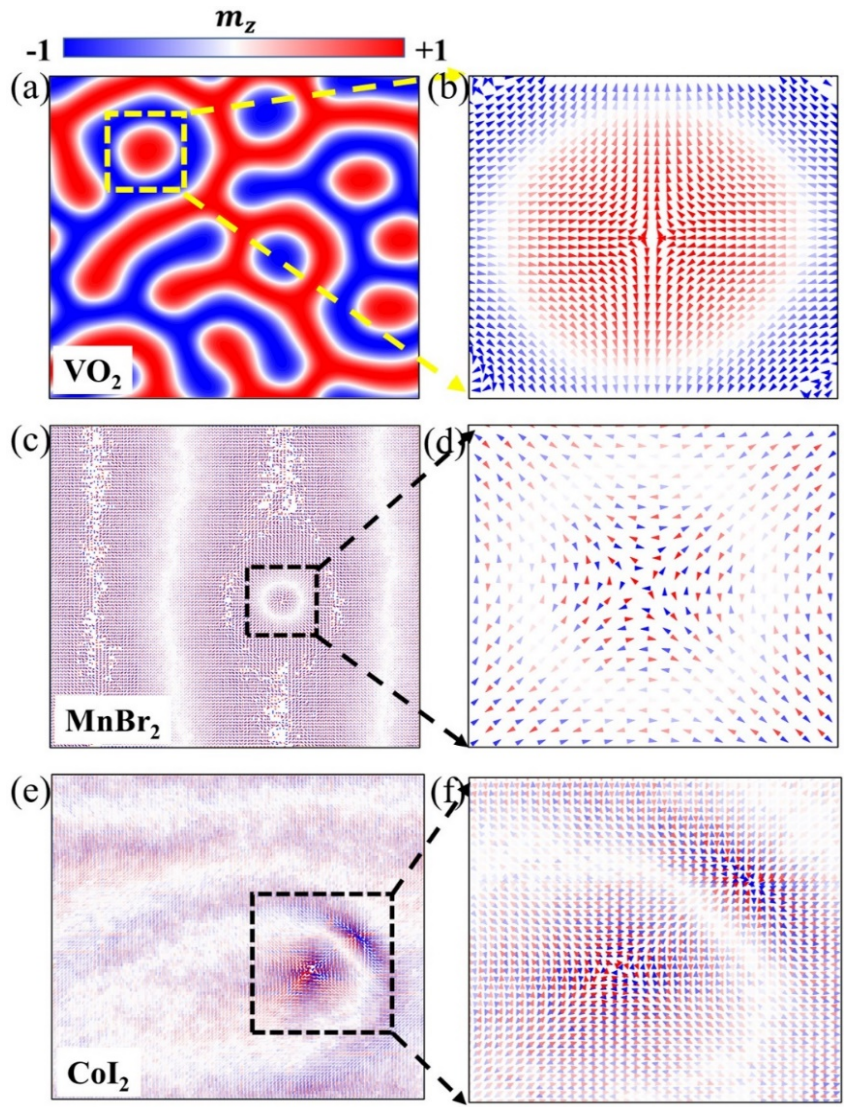

**Figure 5.** Real-space spin configurations obtained from atomistic spin model simulations. Spin textures of a 60×60 nm square and zooms of topological quasiparticle (indicated by the dashed square) of $VO_2$ (**a, b**), $MnBr_2$ (**c, d**) and $CoI_2$ (**e, f**) monolayers. The color map indicates the out-of-plane spin component, and the arrows indicate the orientation of in-plane spin component.